\documentclass[aps,prl,column,showpacs,reprint,superscriptaddress]{revtex4-1}
\usepackage{wrapfig}
\usepackage[]{graphicx,xcolor}
\usepackage{tabularx}
\usepackage{booktabs}
\usepackage{textcomp}
\usepackage{amsmath}
\usepackage{amssymb}
\usepackage[]{graphics}
\usepackage{amssymb}
\usepackage{amsmath}

\begin{document}

\title{Near-Field Radiative Heat Transfer Between Metasurfaces: A Full-Wave Study Based on 2D Grooved Metal Plates }

\author{Jin~Dai}
\affiliation{Department of Materials and Nano Physics, School of Information and Communication Technology, KTH-Royal Institute of Technology, Electrum 229, 16440 Kista, Sweden}
\email[]{e-mail: jind@kth.se}
\author{Sergey~A.~Dyakov}
\affiliation{Photonics and Quantum Material Center, Skolkovo Institute of Science and Technology,
 Nobel street 3, Moscow, Russia}
\author{Sergey~I.~Bozhevolnyi}
\affiliation{Centre for Nano Optics, University of Southern Denmark, Campusvej 55, DK-5230 Odense, Denmark}
\author{Min~Yan}
\affiliation{Department of Materials and Nano Physics, School of Information and Communication Technology, KTH-Royal Institute of Technology, Electrum 229, 16440 Kista, Sweden}
\email[]{e-mail: miya@kth.se}

\date{\today}
\begin{abstract}
Metamaterials possess artificial bulk and surface electromagnetic states. Tamed dispersion properties of surface waves allow one to achieve controllable super-Planckian radiative heat transfer~(RHT) process between two closely spaced objects. We numerically demonstrate enhanced RHT between two 2D grooved metal plates by a full-wave scattering approach. The enhancement originates from both transverse magnetic spoof surface plasmon polaritons and a series of transverse electric bonding- and anti-bonding waveguide modes at surfaces. The RHT spectrum is frequency-selective, and highly geometrically tailorable. Our simulation also reveals thermally excited non-resonant surface waves in constituent materials can play a prevailing role for RHT at an extremely small separation between two plates, rendering metamaterial modes insignificant for the energy transfer process.
\end{abstract}


\maketitle 
Energy can be transferred between two noncontact bodies with different temperatures in a vacuum environment  through thermal radiation. At far-field regime, the energy transferred through thermal radiation is limited by that between two blackbodies. At near-field regime, i.e. when the separation between two bodies is much smaller than the thermal wavelength of the system, photon tunneling of both resonant and non-resonant evanescent waves can make significant contributions to energy transfer~\cite{Polder1971,RevModPhys}. This phenomenon has been widely studied for many different configurations, from simple homogeneous plate-plate structure~\cite{plateFransoeur,Dyakov2014}, sphere-sphere structure~\cite{PRLsphere,PRBsphere1,PRLsphere2}, sphere-plate structure~\cite{sphereplate,Narayanaswamy2008,Rousseau2009exp}, to more delicate periodic nano-structures~\cite{Rodriguez2011,PRBrgrating,Jingrating,Jingrating2}. While most studies focus on numerical investigations, recent technological advances make measurements of radiative heat transfer~(RHT) across a gap size as small as a few nanometers accessible ~\cite{Song2015,Kim2015,Song2016}. The emerging field of near-field RHT opens up possibilities for several promising applications, such as nano-gap thermo-photovoltaics~(TPVs)~\cite{TPV1,TPV2}, thermal memory devices~\cite{memory1,memory2}, radiative cooling~\cite{cool1,cool2}, and super-resolution lithography~\cite{Pendry}, etc. 
\begin{figure}[htb!]
\centering
\includegraphics[width=1\columnwidth]{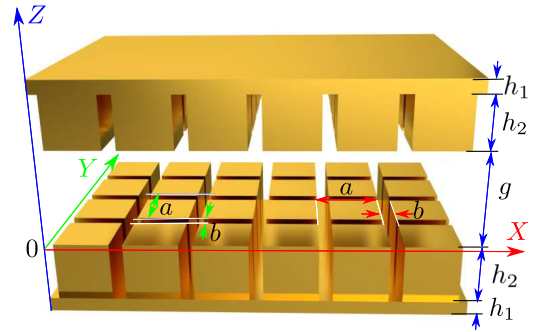}
\caption{Schematic of the proposed two 2D grooved metal metasurfaces  with a square lattice with period of $a=1~\mu$m separated by vacuum gap of $g$. The thickness of homogeneous gold substrate is $h_1=1~\mu$m. The groove depth and groove width are $h_2=5~\mu$m and $b=200$~nm. The temperatures of the bottom- (hot) and top- (cold) plate are $T_1=301$~K and $T_2=300$~K respectively.}
\label{fig1}
\end{figure}

It is known that electromagnetic surface waves, owing to their high spectral density of states, contribute critically to enhanced near-field RHT. Most investigations mentioned above are based on homogeneous natural materials (typically polar dielectrics such as SiC and SiO2, or heavily doped semiconductors), where their intrinsic surface waves are utilized for enhancing RHT in the near-field regime. The lack of dispersion tailorability for surface waves of natural materials severely limits their achieving of spectral controlled RHT for various applications, especially at different temperature settings. Deviating from homogeneous bodies, there have been several studies of RHT between two 1D structured bodies \cite{ Rodriguez2011,PRBrgrating,Jingrating,Jingrating2}. With adequate material and geometry choices, a single 1D structured plate can sustain guided modes below light line; presence of evanescent waves leads to enhanced RHT. However, modes of a 1D structured plate are polarization dependent, and highly anisotropic along different surface-parallel wave vector directions. Here, we examine near-field RHT between two 2D grooved metal plates~(GMPs), which support artificial surface waves of similar dispersion properties along all surface-parallel directions. Besides transverse magnetic~(TM) spoof surface plasmon polariton~(SSPP) modes, which are physically the same as SSPP modes supported by 1D GMPs~\cite{Jingrating,Jingrating2}, there exists a series of bonding and anti-bonding guided transverse electric~(TE) mode pairs contributing to near-field energy transfer. The resulted near-field transmission factor is almost perfectly isotropic along different surface-parallel directions, a phenomenon one can expect for RHT between two homogenous plates made of natural material, e.g. SiC. We study the dependence of near-field RHT on plate separation ranging from a few micrometers to sub-100 nanometers. At extremely near-field limit, the RHT is found to be dominantly contributed by the non-resonant evanescent waves due to frustrated total internal reflection. At the same time, the Derjaguin’s proximity approximation~(PA) ~\cite{PRBrgrating1} starts to take effect; that is, RHT between two 2D GMPs tends to converge to RHT between two homogeneous metal plates weighted by the metal filling factor. Similar tendency was noted in Ref.~[\onlinecite{Rodriguez2011}], based however on a single-polarization, single-wave-vector-plane analysis. Our finding shows in a more general scenario that geometrically structuring cannot increase RHT between two homogeneous plates at extremely near-field limit. Although our numerical investigation is based on 2D GMPs, the phenomena observed should share similar features with other metasurface designs. In fact, the 2D GMP is arguably one of the simplest isotropic metasurface designs with tailorable surface wave dispersions that one can realize experimentally.

The schematic of the structure studied is shown in Fig.~\ref {fig1}.  It consists of two identical 2D periodically grooved gold plates separated by a vacuum gap $g$. Without loss of generality, we choose a simple square lattice in our simulations. The period $a=1~\mu$m, width of grooves $b=200$~nm, depth of grooves $h_2=4.7~\mu$m, and the thickness of the homogeneous gold substrate $h_1=1~\mu$m are fixed. The temperatures of the bottom~(hot) and top~(cold) plate are $T_1=$301~K and $T_2=$300~K, respectively. The optical constant of gold is described by a Drude mode, $\epsilon_{\mathrm{Au}}(\omega)=1-\frac{\omega_p^2}{\omega(\omega+i\gamma)}$, in which $\omega_p=9$~eV, and $\gamma=35$~meV.

The radiative heat flux between two 2D periodic metasurfaces with a square lattice can be described by
\begin{widetext}
\begin{align}
\label{eq1}
q(T_1, T_2)=\frac{1}{8\pi^3}\int_0^\infty[\Theta(\omega, T_1)-\Theta(\omega, T_2)] \mathrm{d}\omega \newline
\int_{-\frac{\pi}{a}}^{+\frac{\pi}{a}}\int_{-\frac{\pi}{a}}^{+\frac{\pi}{a}}\Phi_{12}(\omega,k_x,k_y)\mathrm{d} k_x\mathrm{d} k_y,
\end{align}
\end{widetext}
in which  $\Theta(\omega, T)$=$\hbar\omega/$exp$[(\hbar\omega/k_B T)-1]$ is the mean energy of Planck oscillators at temperature $T$ and angular frequency $\omega$. $\Phi_{12}(\omega,k_x,k_y)$ is the transmission factor that describes the probability of two thermally excited photons, one $\mathbf{s}$-polarized~(TE) and one $\mathbf{p}$-polarized~(TM), with surface-parallel wavevector ($k_x$,$k_y$) at angular frequency $\omega$ transferring from one metasurface to the other. The transmission factor in the scattering approach can be written as
\begin{subequations}
\label{eq3}
\begin{eqnarray}
\Phi(\omega,k_x,k_y) &=& \mathrm{Tr}(\mathbf{DW_{1}D^{\dagger}W_{2}}),\label{equationa}
\\
\mathbf{D} &=& \mathbf{(1-S_{1}S_{2})^{-1}}, \label{equationb}
\\
\mathbf{W_{1}} &= &\mathbf{\sum_{-1}^{pw}-S_{1}\sum_{-1}^{pw}S_{1}^{\dagger}+S_{1}\sum_{-1}^{ew}-\sum_{-1}^{ew}S_{1}^{\dagger}}, \label{equationc}
\\
\mathbf{W_{2}} &= &\mathbf{\sum_{1}^{pw}-S_{2}^{\dagger}\sum_{1}^{pw}S_{2}+S_{2}^{\dagger}\sum_{1}^{ew}-\sum_{1}^{ew}S_{2}}, \label{equationd}
\\
\mathbf{S_{1}} &= &\mathbf{R_{1}(k},\omega), \label{equatione}
\\
\mathbf{S_{2}} &= &e^{ik_{z}g}\mathbf{R_{2}(k},\omega)e^{ik_{z}g}. \label{equationf}
\end{eqnarray}
\end{subequations}
Here, $\mathbf{R_{1}}$ and $\mathbf{R_{2}}$ are the reflection operators of the bottom and  top metasurfaces respectively. The operators $\mathbf{\sum_{-1(+1)}^{pw(ew)}}$ are used for identifying propagating and evanescent waves~\cite{nummethod}. Since the two meatasufaces are identical for the configuration under study, $\mathbf{R_{1}}$ is equal to $\mathbf{R_{2}}$. For homogeneous parallel-plate structure they are simplified to the scalar Fresnel coefficients. However, for nanostructured metasurfaces the reflection operators have to take non-specular diffractions into consideration. To calculate these reflection operators for metasurfaces, we use the rigorous coupled wave analysis (RCWA)~\cite{Li}. It is noteworthy that we have implemented Li's factorization rule, which applies inverse rule in one direction and Laurent's rule in the other direction, to improve the convergence for metals~\cite{Li}.
\begin{figure}[htb!]
\centering
\includegraphics[width=1\columnwidth]{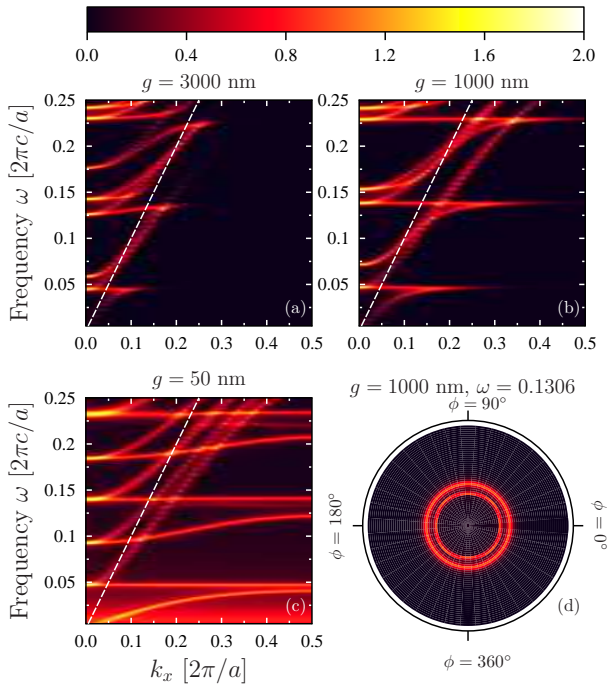}
\caption{Transmission factors $\Phi(\omega,k_x,0)$  between two grooved metal metasufaces with in-plane wavevector along $\Gamma$-$X$ direction as a function of angular frequency $\omega$ and $k_x$ for gap size of (a)~g=$3000$~nm, (b)~g=$1000$~nm, and (c)~g=$50$~nm. (d) Transmission factor at frequency $\omega=0.1306~[2\pi c/a]$ for the same structure as in (b) but for surface-parallel vector directions ranging from 0$^{\circ}$ to 360$^{\circ}$, with respect to $+x$ axis.}
\label{fig2}
\end{figure}

To investigate the contribution of thermally excited photons at different states to RHT, we plot transmission factor $\Phi(\omega,k_x,0)$ along $\Gamma$-$X$ direction as a function of angular frequency $\omega$ in Fig.~\ref {fig2}~(a-c) for three scenarios with different gap sizes of 3000, 1000, and 50~nm, respectively. The high-transmission states form clearly dispersion curves of the thermally excited modes that are involved in the heat transfer process~\cite{Jingrating,Jingrating2}.  At $k_x=0$, $k_y=0$, the $\mathbf{s}$- and $\mathbf{p}$- polarization are degenerate due to  $\mathbf{C4}$ symmetry of the GMP. The TM-polarized SSPP modes, as represented by the relatively flat bands in the plots, shares the same physical origin as they are in 1D GMPs~\cite{Jingrating,Jingrating2}, but exsiting now with all surface-parallel wavevector directions. Besides the SSPP modes, there exist a series of bonding and anti-bonding TE guided mode pairs. Both categories of modes will be clarified in more detail as we present their precise dispersions in the next figure. As the gap decreases, the contribution from surface modes, which stay below the light line, becomes more significant, suggesting further enhanced RHT. For gap size as small as 50~nm, modes with $\mathbf{k}$-vector as large as $k_x=\pi/a$, which is at the first Brillouin zone boundary, can even contribute to the RHT process, which nearly maximized RHT at those resonance frequencies. Figure.~\ref{fig2}(c) also reveals significant transmission contribution from a continuum of states at low frequencies, which is due to tunneling of non-resonant $\mathbf{s}$-polarized evanescent waves through frustrated total internal reflection~\cite{tir}. It is worth mentioning that, at even smaller gap size (sub-nanometer), the $\mathbf{p}$-polarized photons may start playing an important role; however non-local effects have to be then considered~\cite{tir}, which is out of the scope of the current study. In Fig.~\ref{fig2}(d) we show the transmission factors at $\omega$=0.1306~[$2\pi c/a$] for surface-parallel $\mathbf{k}$-vectors with different directions from 0 to 360 degrees, for the $g$=1000~nm configuration. The transmission factor along different surface-parallel wave vector directions exhibits good uniformity. Such isotropic transmission factor is preserved even at SSPP-resonance frequencies, i.e. flat-band positions in Figs.~\ref{fig2}(a-c). This gives rise to a better frequency-selectivity in spectral heat flux, which one obtains by integrating transmission factor over all possible surface-parallel wave vectors.
\begin{figure}[htb!]
\centering
\includegraphics[width=1\columnwidth]{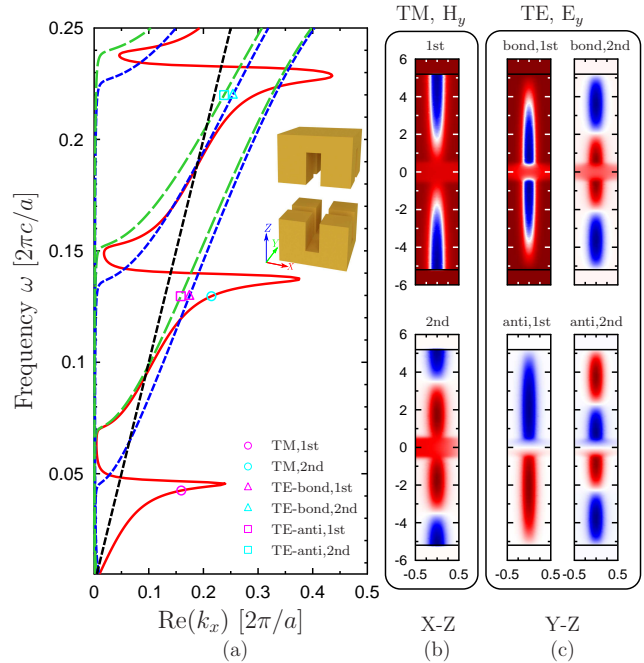}
\caption{(Color online) The relationship between real part of complex wavevector ${k_x}$ and angular frequency for two 2D GMPs separated by a vacuum gap of $g=1000$~nm. The solid red curve denote the TM modes, and the dashed lines denote the bonding~(blue) and anti-bonding~(green) TE modes. The black dashed line indicates the light line in vacuum. The markers denote the mode positions on the band diagram whose corresponding field distributions are mapped on the right.}
\label{fig3}
\end{figure}

In order to illustrate the physical mechanism of these modes, we use a finite element method to calculate the complex $\mathbf{k}$ vs. $\omega$ dispersion relation along $\mathbf{\Gamma}$-$\mathbf{X}$ for two GMPs separated by a vacuum gap of $g=1000$~nm. We used a unit cell depicted in the inset of Fig.~\ref{fig3}(a); periodic boundary condition are imposed on four side truncation planes. The calculated real part of $k_x$ as a function of angular frequency $\omega$ is presented in Fig.~\ref {fig3}(a) There are two types of modes in general, the curve containing flat portions is due to TM-polarized SSPP modes; the more dispersive bands are due to TE-polarized guided modes. To illustrate more clearly, we visualize in Figs.~\ref{fig3}(b-c) the TM and TE modes at specific frequencies as indicated by markers in panel (a). For TM modes, the $y$-component of magnetic field $H_y$ in $x$-$z$ plane (across the unit cell center) are plotted. For TE modes, the $y$-component of electric fields $E_y$ in $y$-$z$ plane is plotted. The TM modes originate from coupling between thermally excited SSPP waves in two plates owing to presence of grooves oriented along $y$ direction~\cite{Jingrating,Jingrating2}. The TE modes, new in this study, originate from interplay of thermally excited guided modes by each of the $x$-oriented grooves. A guided TE mode exists due to the vacuum-gap interface serves as approximately a perfect magnetic conductor and the groove bottom serves as a perfect electric conductor. When two grooved plates are brought close to each other, mode coupling splits each TE band into bonding and anti-bonding bands in the dispersion curves. Contributions of both TE and TM modes to the RHT process can be clearly identified as we compare their dispersion curves in Fig.~\ref{fig3}(a) and the transmission factor map plotted in Fig.~\ref{fig2}(b). In Fig.~\ref{fig2}(b), the intersections of TE and TM bands imply possibility of degenerate states of two polarizations, corresponding to a maximum value of 2 in transmission factor. The TE modes exhibit cutoff decided by the groove depth. They always exist for wavelengths smaller than four times of the groove depth; for the groove depth in our case study, consideration of thermal excited TE modes is indispensable.
\begin{figure}[htb!]
\centering
\includegraphics[width=1\columnwidth]{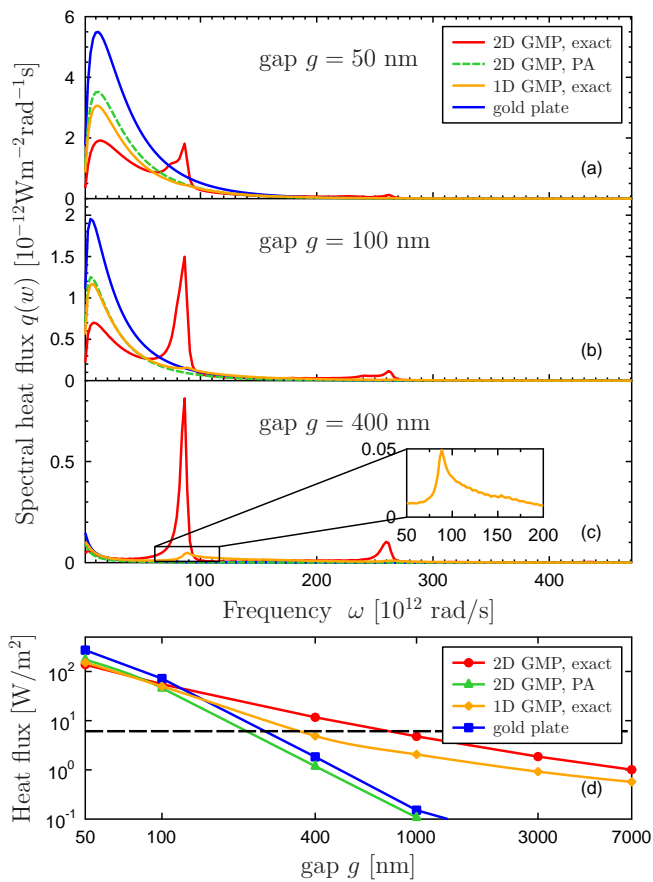}
\caption{(Color online) Spectral heat flux between two 2D grooved metal plates with exact solution using a scattering approach~(solid red line) and PA~(dashed red line), two 1D grooved metal metasurfaces, and two $5.7~\mu$m gold plates for gap size of (a)~$g=400$~nm, (b)~$g=100$~nm, (c)~$g=50$~nm. (d)~The integrated heat flux between these four configurations as a function of gap size. The black dashed line denotes the heat transfer between two blackbodies.}
\label{fig4}
\end{figure}

In Figs.~\ref{fig4}(a-c) we show numerically calculated radiative heat flux spectra between two 2D GMPs. In the plots, we also superimpose flux spectrum between two 1D GMPs, and  that between two homogeneous gold plates of thickness $5.7~\mu$m. Panels (a)-(c) are for three different gap sizes. The period and the geometry of the grooves of 1D GMPs are identical to those of 2D GMPs. It is seen that for gap size $g=400~$nm (or larger, not shown), heat transfer is mainly through the first-order of SSPP channel, resulting in a sharp peak at angular frequency around $\omega=86\times 10^{12}$~rad/s. The peak at $\omega=262\times 10^{12}$~rad/s, which has a lower amplitude due to the energy distribution of Planck oscillators, comes from the contribution of the second-order SSPPs. Note that, if one considers higher temperatures of the GMPs, the contribution from the second- or even higher-order SSPPs may play a more important role in RHT. Since the TE modes exhibit very dispersive profiles, their contribution is broadband compared to the contribution from SSPP modes. As the gap size decreases, a broad peak at very low frequencies appears and dominates the heat transfer at a gap size of  50~nm. This peak has a similar profile as that for homogeneous gold plates. It is attributed to the tunneling of thermally excited $\mathbf{s}$-polarized photons though frustrated total internal reflection~\cite{tir}. The fact that the broad peak weighs much more than other peaks due to geometry-induced surface waves motivates us to check the validity of Derjaguin’s proximity approximation~(PA). It turns out that PA can capture the basic feature, at the low-frequency regime, of the radiative heat flux in Figs.~\ref{fig4}(a-b). It suggests tunneling of non-resonant evanescent waves plays a key role in RHT at very small gap distances. Overall, at a moderate gap size~(e.g.~400~nm), RHT between two 2D GMPs can be much higher than that between two 1D GMPs.

An important spectral feature of using 2D GMPs in RHT is that SSPP resonances give rise to sharper and higher peaks in heat-flux spectrum, compared to using 1D GMPs. It is owing to the isotropic dispersion properties of SSPP modes supported by the 2D GMP, examined from the surface-parallel wave vector directions. Such superior frequency selectivity in RHT can be attractive for, e.g. near-field TPV applications.

In Fig.~\ref{fig4}(d) we plot integrated heat flux for the structures demonstrated in panels (a-c) as a function of gap size. It shows that for gap size larger than 100~nm GMPs can enhance RHT compared to two homogeneous gold plates. The enhancement is close to two orders in magnitude when gap size is at 1~$\mu$m, and increasing towards larger gap sizes. However, for sub-100~nm gap sizes, tunneling of non-resonant evanescent waves prevails over tunneling of geometry-induced TM/TE modes; the homogeneous-plate configuration has the highest RHT but poor frequency selectivity.

In conclusion, through full-wave numerical calculations, we have thoroughly examined RHT between two 2D periodic GMPs separated by a gap size ranging from a few micrometers down to a few tens of nanometers. Structure-induced TM SSPP modes and TE guided modes, whose frequencies can be geometrically tailored, are found to contribute to much enhanced RHT, compared to two homogeneous plates. The isotropic TM SSPP dispersion properties along different surface-parallel directions lead to superior frequency selectivity and leveraged peak values in RHT. The spectral tailorability, frequency selectivity, and much enhanced total or peak RHTs are typical advantages of using artificial metasurfaces in advanced RHT management. Tamed RHT can be technologically important for applications like near-field TPV, super-resolution heating-based lithography, thermal logic and memory devices, etc. Our simulation also reveals that at extremely small gap sizes, metamaterial modes due to introduction of fine geometry may not positively contribute to RHT, as then the dominant RHT process is tunneling of non-resonant evanescent waves.
\begin{acknowledgments}
J. D., and M. Y. acknowledge the support by the Swedish Research Council (Vetenskapsr\aa det or VR) via Project No. 621-2011-4526, and VR's Linnaeus center in Advanced Optics and Photonics (ADOPT). The simulations were performed on resources provided by the Swedish National Infrastructure for Computing (SNIC) at PDC Centre for High Performance Computing (PDC-HPC)
\end{acknowledgments}
%
\bibliographystyle{apsrev4-1}

\end{document}